\documentclass[twocolumn,nofootinbib,superscriptaddress,showpacs]{revtex4-1}

\usepackage{hyperref}
\usepackage{amsmath,amssymb,amsfonts,amsthm,latexsym,stmaryrd}

\def\be{\begin{equation}}
\def\ee{\end{equation}}
\def\ba{\begin{eqnarray}}
\def\ea{\end{eqnarray}}
\def\bas{\begin{subequations}\begin{eqnarray}}
\def\eas{\end{eqnarray}\end{subequations}}

\def\eps{\varepsilon}
\def\lp{\ell_\text{Pl}}
\def\Tr{\text{Tr}}

\def\la{\langle}
\def\ra{\rangle}
\def\de{\mathrm{d}}

\def\f{\frac}

\def\SU{\text{SU}}

\def\SL{\text{SL}}
\def\su{\mathfrak{su}}

\def\sll{\mathfrak{sl}}

\def\i{\mathrm{i}}
\def\nn{\nonumber}
\def\ds{\displaystyle}

\usepackage{titlesec}
\titleformat*{\section}{\small\center\bfseries}

\newenvironment{flist}{
\begin{list}{\labelitemi}{\leftmargin=1.3em}
}{\end{list}}

\begin{document}


\title{Near-Horizon Radiation and Self-Dual Loop Quantum Gravity}

\author{Marc Geiller}
\email{mgeiller@gravity.psu.edu}
\affiliation{Institute for Gravitation and the Cosmos \& Physics Department, Penn State, University Park, PA 16802, U.S.A.}
\author{Karim Noui}
\email{karim.noui@lmpt.univ-tours.fr}
\affiliation{Laboratoire de Math\'ematiques et Physique Th\'eorique, Universit\'e Fran\c cois Rabelais, Parc de Grandmont, 37200 Tours, France}
\affiliation{Laboratoire APC -- Astroparticule et Cosmologie, Universit\'e Paris Diderot Paris 7, 75013 Paris, France}

\begin{abstract}
We compute the near-horizon radiation of quantum black holes in the context of self-dual loop quantum gravity. For this, we first use the unitary spinor basis of $\text{SL}(2,\mathbb{C})$ to decompose states of Lorentzian spin foam models into their self-dual and anti self-dual parts, and show that the reduced density matrix obtained by tracing over one chiral component describes a thermal state at Unruh temperature. Then, we show that the analytically-continued dimension of the $\text{SU}(2)$ Chern--Simons Hilbert space, which reproduces the Bekenstein--Hawking entropy in the large spin limit in agreement with the large spin effective action, takes the form of a partition function for states thermalized at Unruh temperature, with discrete energy levels given by the near-horizon energy of Frodden--Gosh--Perez, and with a degenerate ground state which is holographic and responsible for the entropy.
\end{abstract}

\maketitle

\section{Introduction}
\label{sec:intro}

\noindent The celebrated work of Bekenstein and Hawking \cite{BHentropy} has shown that black holes exhibit thermodynamical properties and can be assigned a temperature as well as an entropy equal to one fourth of their horizon area. It is generally expected from the candidate theories of quantum gravity to provide a framework for understanding the origin of this thermodynamical behavior, although arguably hard to define more precisely what these expectations should be.

From the point of view of loop quantum gravity (LQG) \cite{LQG}, the derivation of black hole entropy relies essentially on the idea that the macroscopic horizon area is realized as the sum of microscopic contributions (quanta of area) excited by the spin network edges that puncture the horizon \cite{BHentropyLQG}. In this picture, the counting of the microstates leads to the Bekenstein--Hawking area law provided that a free parameter of the theory that enters the area spectrum, the Barbero--Immirzi parameter $\gamma$, is fixed to a particular finite and real value. The fact that this parameter seems to play such a crucial role in the quantum theory even though it is totally irrelevant at the classical level has been at the center of important debates \cite{gamma}.

In any approach to the derivation of black hole entropy from a more fundamental theory (as opposed to semiclassical gravity), the various constants entering the definition of this latter are expected to run between the UV and the IR, until they reach a certain ``renormalized'' value in the effective theory. Only then can one compare the result of an entropy calculation with the semiclassical answer of Bekenstein and Hawking. From this perspective, the very idea of fixing $\gamma$ to a particular value can seem premature. This is especially true in the absence of an understanding of the full dynamics of LQG and of its continuum limit.

In spite of this observation, it might still be the case that working with the imaginary value $\gamma=\i$ is the only choice which will lead to a consistent quantum theory, accompanied by a proper (canonical or covariant) dynamics and possessing a correct semiclassical and continuum limit. This expectation is of course motivated by the geometrical significance carried by the self-dual Ashtekar connection defined for $\gamma=\i$. While quantizing the self-dual theory remains an immensely hard problem because of group-theoretical difficulties, one reasonable point of view is to see $\gamma$ as a regulator whose presence is reminiscent of the fact that LQG is constructed in terms of the subgroup $\SU(2)$ of $\SL(2,\mathbb{C})$. Having in mind the idea of a ``generalized Wick rotation'', a natural question to ask is wether it is possible to take advantage of the presence of $\gamma$ in the quantum theory in order to analytically continue relevant mathematical structures to their analogues in the full Lorentz group.

This idea was applied in \cite{FGNP} in the context of the $\SU(2)$ Chern--Simons theory description of black holes in LQG (see also \cite{Smolin} for earlier insights). It was shown that for a fixed number of punctures and in the limit of large spins, the analytically-continued dimension of the $\SU(2)$ Chern--Simons Hilbert space (which can be understood as the weight assigned to a black hole configuration) behaves as\footnote{We work in units in which $k_\text{B}=c=1$. The Planck length is given by $\lp=\sqrt{G\hbar}$, where $G$ is Newton's constant.} $\exp(A/4\lp^2)$. Interestingly, it was observed that the same analytically-continued dimension arises naturally in the description of the three-dimensional BTZ black hole \cite{BTZ} (in this case however, it is the sign of the cosmological constant and not $\gamma$ that plays the role of a regulator).

The proposal of \cite{FGNP} is supported by remarkable additional facts. First, it was shown in \cite{KMS} that it is necessary to work with $\gamma=\i$ in order for the horizon states to posses a geometrical KMS interpretation (see also \cite{DanieleCFT} for the relationship with conformal field theory). Second, the work \cite{CMC} reveals that when working in the constant mean curvature gauge and in the presence of a conformally-coupled scalar field, it is not possible to fix $\gamma$ to a particular real value (since this value would then depend on the mass of the black hole), and only the analytic continuation to $\gamma=\i$ yields the correct entropy formula. Finally, building up on the fact that the on-shell action for Lorentzian gravity with a suitable boundary term possesses an imaginary part \cite{Yasha}, it was noted in \cite{BN} that the effective action obtained from the large spin limit of spin foam amplitudes is compatible with this observation if and only if one chooses the self-dual value of $\gamma$ after the computation of the asymptotics\footnote{Defining the spin foam amplitudes themselves with $\gamma=\i$ is still an open issue. If this turns out to be possible, then the analysis of the large spin limit would need to be studied again.}. This reveals for the first time that there is an agreement between the derivation of the entropy and that of an effective action in the regime of large spins (which is coined ``transplanckian'' in \cite{BN}).

Although the perturbative study of Euclidean quantum gravity exhibits a UV fixed point at the Euclidean self-dual value $\gamma=\pm1$ \cite{Benedetti-Speziale}, the question of wether the Lorentzian parameter will run to $\pm\i$ is still open (within any renormalization scheme). We therefore leave this aspect aside, and embark on studying further the proposal made in \cite{FGNP} for describing black hole thermodynamics within the self-dual quantum theory. We hope that this will motivate and guide the developement of a full consistent picture.

As an additional motivation, we would like to clarify the relationship between black hole thermodynamics in the self-dual quantum theory and the results of Bianchi \cite{BianchiBH}. In this work, it was shown that the near-horizon thermodynamical properties of black holes can be reproduced using ingredients from spin foam models, and in particular the idea that the horizon energy is described by the generator of the boost Hamiltonian acting on states of the EPRL/FK model \cite{EPRL/FK}. As shown by Bianchi, this input is sufficient in order to derive the fact that the energy and the temperature of the quantum horizon are given respectively by the Frodden--Gosh--Perez energy \cite{FGP} and the Unruh temperature \cite{UnruhT}. Using the Clausius relation relating these two quantities to the entropy, one then obtains the entropy of non-extremal black holes independently of the value of $\gamma$ (at leading order in the area).

After recalling in section \ref{sec:2} the essential ingredients of the near-horizon description of \cite{FGP}, we are going to re-derive in section \ref{sec:3} the thermal behavior of the quantum horizon. However, while \cite{BianchiBH} probes the temperature of the horizon using a thermometer, we are going to construct a density matrix by using the so-called unitary spinor basis of $\SL(2,\mathbb{C})$, and then exploit the explicit separation between self-dual and anti self-dual representations provided by this basis in order to trace out one chiral component. We interpret the resulting thermal behavior as an indication of the fact that the fundamental excitations of quantum geometry near the horizon are indeed described by self-dual representations. Finally, in section \ref{sec:4} we will discuss the thermodynamical properties which seem to be encoded (in addition to the entropy) in the analytically-continued Chern--Simons dimension of \cite{FGNP}.

\section{The near-horizon picture}
\label{sec:2}

\noindent The difficulty with quantum gravity derivations of black hole entropy is to identify what is actually meant by a black hole, and which ingredients of the computation know about these properties. As we will recall in the next section, computations in the canonical framework rely on the notion of isolated horizons.

The analysis of \cite{BianchiBH} is carried out from the perspective of a stationary observer at proper distance $l=a^{-1}$ from the horizon, where $a$ is the acceleration. As shown in \cite{FGP}, there is a well-defined notion of energy associated to such an observer, which is given by
\be\label{FGP energy}
E=\f{A}{8\pi G}a,
\ee
where $A$ is the horizon area. In this setup, there is literally no difference between Hawking radiation and the Unruh effect, and one can focus on the Rindler horizon seen by the observer. As argued in \cite{BianchiBH}, the Hamiltonian operator given by the generator of Lorentz boosts times the acceleration possesses two noticeable properties. First, it generates the evolution along uniformly accelerated trajectories, and therefore the evolution in the presence of the Rindler horizon which is seen by the observer outside of the black hole. Second, it measures the energy of the Rindler horizon. This can indeed be seen by evaluating the Hamiltonian on the so-called $\gamma$-simple states of the EPRL/FK spin foam models which color a quantum element of surface, and noticing that the expectation value reproduces the classical near-horizon energy \eqref{FGP energy}.

We therefore have two ingredients. The first one, ensuring that we are describing the physics of a Rindler observer, is that the energy is labelled by eigenstates of the boost generator. The second one, relating this description with quantum gravity, is that the states themselves are labelled by unitary representations of the Lorentz group compatible with the covariant spin foam dynamics of LQG.

We are now ready to decompose these states into (anti) self-dual parts, and to construct the reduced density matrix.

\section{Self-dual decomposition of states and thermality}
\label{sec:3}

\noindent Instead of decomposing the representations of $\SL(2,\mathbb{C})$ in terms of that of the subgroup $\SU(2)$ as recalled in appendix \ref{appendix1}, let us switch instead to the so-called unitary spinor (or helicity) basis \cite{Helicity1,Huszar}. For this, we first introduce the usual (anti) self-dual generators $\vec{J}^\pm=(\vec{L}\pm\i\vec{K})/2$. A natural basis of states associated to this decomposition is given by the vectors $|(\rho,k),(\mu,\nu)\ra$, and denoted by $|\mu,\nu\ra$ for short. These vectors diagonalize the third component of the rotation and boost generators according to
\be\nn
L_3|\mu,\nu\ra=\mu|\mu,\nu\ra,\qquad K_3|\mu,\nu\ra=\nu|\mu,\nu\ra,
\ee
where $\mu\in\mathbb{N}/2$ and $\nu\in\mathbb{R}$. The eigenvalues of $J^\pm_3$ are therefore given by $m_\pm=(\mu\pm\i\nu)/2$, and one can furthermore introduce the complex spins $\ell_\pm=(k\pm\i\rho-1)/2$ defined in such a way that the Casimir $(\vec{J}^\pm)^2$ is given by $\ell_\pm(\ell_\pm+1)$. With these notations, the vectors $|\mu,\nu\ra$ can be equivalently written as $|m_+(\mu,\nu)\ra\otimes|m_-(\mu,\nu)\ra$, which makes explicit the (anti) self-dual decomposition $\sll(2,\mathbb{C})=\su(2)_\mathbb{C}\oplus\su(2)_\mathbb{C}$.

The change of basis is given by the relations
\ba\label{basis change}
|\mu,\nu\ra&=&\sum_{j,n}\la j,n|\mu,\nu\ra|j,n\ra,\nn\\
|j,n\ra&=&\int_{-\infty}^{+\infty}\de\nu\sum_\mu\la\mu,\nu|j,n\ra|\mu,\nu\ra,
\ea
and the overlap coefficients $\la j,n|\mu,\nu\ra=\overline{\la\mu,\nu|j,n\ra}$ have been computed in \cite{Huszar} (the generic expression is recalled in appendix \ref{appendix2}).

We now want to apply this change of basis to the states that are used to describe the covariant spin foam dynamics of the quantum theory. In the case of the EPRL/FK model, these are labelled by $\gamma$-simple representations of the type $|(p,k),(j,n)\ra=|(\gamma(j+1),j),(j,n)\ra$. The state of an elementary quanta of surface with normal in the third direction is given by the highest weight element $|(\gamma(j+1),j),(j,j)\ra$ which we will simply denote by $|0(j)\ra$. This notation is reminiscent of a vacuum as described by the Rindler observer.

Using the formulas recalled in appendix \ref{appendix2}, one can show that the overlap coefficient (which we square for convenience) between $|0(j)\ra$ and the helicity basis are given by
\ba\label{overlap}
&&\big|\la\mu,\nu|0(j)\ra\big|^2\\
&=&
\left\{
\begin{array}{l}
\delta_{\mu,j}\f{\ds\cosh(\pi\rho)}{\ds\sinh(\pi\rho)+\sinh(\pi\nu)}P_j(\rho,\nu)
\phantom{\delta_{\mu,j}\f{\ds\sinh(\pi\rho)}{\ds\cosh(\pi\rho)+\cosh(\pi\nu)}Q_j(\nu)}
\hspace{-4.2cm}\text{if }j\text{ half-integer},\vspace{0.3cm}\\
\delta_{\mu,j}\f{\ds\sinh(\pi\rho)}{\ds\cosh(\pi\rho)+\cosh(\pi\nu)}Q_j(\rho,\nu)
\phantom{\delta_{\mu,j}\f{\ds\cosh(\pi\rho)}{\ds\sinh(\pi\rho)+\sinh(\pi\nu)}P_j(\nu)}
\hspace{-4.2cm}\text{if }j\text{ integer},
\end{array}
\right.\nn
\ea
where
\be\nn
P_j(\rho,\nu)=\f{(2j+1)(\rho+\nu)}{\left(2j+1\right)^2+(\rho+\nu)^2}\prod_{i=1}^{j+1/2}\f{4i^2+(\rho+\nu)^2}{(2i-1)^2+4\rho^2},
\ee
and
\be\nn
Q_j(\rho,\nu)=\f{2j+1}{2\rho}\prod_{i=1}^j\f{(2i-1)^2+(\rho+\nu)^2}{4(i^2+\rho^2)}.
\ee
We see that the two results are qualitatively the same, in the sense that the square of the overlap coefficient is given by the product between a hyperbolic function and a rational function.

We can now characterize a pure state with the density matrix
\be\nn
\hat{\rho}=|0(j)\ra\la0(j)|,
\ee
and write this latter in the decoupled basis $|\mu,\nu\ra=|m_+(\mu,\nu)\ra\otimes|m_-(\mu,\nu)\ra$. In terms of the overlap coefficients which we have just computed, this rewriting is given by
\ba
&&\hat{\rho}=\nn\\
&&\int\de m\,\la\mu,\nu|0(j)\ra|m_+\ra\otimes|m_-\ra\la m_+'|\otimes\la m_-'|\la0(j)|\mu',\nu'\ra,\nn
\ea
where the measure is $\de m=\de m_\pm^{\vphantom{'}}\de m_\pm'$, and $\de m_\pm$ is given by an integral over $\nu$ and a sum over $\mu$ as in \eqref{basis change}. Now, following the idea that the observer near the horizon is only sensitive to the self-dual part of this state (or to ``half of the information''), one can trace over the set of anti self-dual representations to obtain the reduced density matrix
\be\label{thermal state}
\hat{\rho}_+=\Tr(\hat{\rho}_-)=\int_{-\infty}^{+\infty}\de\nu\,P(j,\nu)|m_+(j,\nu)\ra\la m_+(j,\nu)|,
\ee
where we have used the fact that the sum over $\mu$ is vanishing if $\mu\neq j$, introduced $P(j,\nu)=\big|\la j,\nu|0(j)\ra\big|^2$,  and $m_+(j,\nu)=(j+\i\nu)/2$.

Now, we would like to propose a physical interpretation of this result in terms of thermalization. From the mathematical point of view, $|m_+(j,\nu)\ra$ are vectors in an infinite dimensional representation of the self-dual part of the Lorentz group. Following the idea of \cite{FGNP} that black holes in LQG should be described by such representations, it is natural to see these as the quantum gravitational excitations seen by a near-horizon observer. Furthermore, as recalled in the previous section, 
such an observer associates to each excitation $|\mu,\nu\ra$ of the Lorentz group a (total) positive energy $E_\text{t}(\nu)$ given by the boost operator (the third component $K_3$ of the boost generator) according to
\be\label{hamiltonian}
H|\mu,\nu\ra=aK_3|\mu,\nu\ra=\text{sign}(\nu)E_\text{t}(\nu)|\mu,\nu\ra,
\ee
where $E_\text{t}(\nu)=a|\nu|$ and $\text{sign}(x)$ is the sign function. As explained above, the excitation $|\mu,\nu\ra=|m_+(\mu,\nu)\ra\otimes|m_-(\mu,\nu)\ra$ does however decompose in terms of its self-dual and anti self-dual components, so it is natural to expect each component of this decomposition to have an energy contribution of $E(\nu)=E_\text{t}(\nu)/2$.

These are the reasons for interpreting \eqref{thermal state} as the quantum gravity realization of a thermal state at Unruh temperature. In fact, when the energy $E(\nu)$ is large, $\nu$ is large independently of the value of $j$. In this regime, it is immediate to show that
\be\nn
P(j,\nu)=N\nu^{2j}e^{-\pi|\nu|}\big(1+\circ(1)\big)=N\nu^{2j}e^{-\beta E(\nu)}\big(1+\circ(1)\big),
\ee
where $N$ is a normalization factor independent of $\nu$, $\beta=2\pi/a$ is the inverse Unruh temperature, and $E(\nu)$ is the energy of the excitation as defined above. This is in agreement with the result of \cite{BianchiBH}, although slightly different in spirit. This result does also hold if one considers the states of the Barrett--Crane model, for which the overlap coefficients are given in appendix \ref{appendix3}. In fact, as shown in appendix \ref{appendix4}, the overlap coefficients exhibit the same asymptotic behavior regardless of the particular states that are considered, which provides the model with a certain robustness.

So far we have considered the regime in which the energy $\nu$ is large and the spin $j$ (or equivalently the area of an elementary surface element) fixed. However, as we will see in the next section, the regime in which one recovers the entropy from the candidate self-dual quantum theory is that of large spins (which as recalled in the introduction corresponds to the only known ``semiclassical'' regime of spin foams). It is therefore natural to consider the behavior of the squared overlap coefficients in the scaling limit in which $\nu$ and $j$ tend to infinity while the ratio $\nu/j$ is kept fixed. In this case, it is immediate to see from \eqref{overlap} that the overlap coefficients behave as:
\be\nn
\big|\la\mu,\nu|0(j)\ra\big|^2\approx\f{g_j(\nu)}{1+e^{\pi(\nu-\rho)}}=\f{g_j(\nu)}{1+e^{\beta E'(\rho,\nu)}},
\ee
where $g_j(\nu)$ is a normalization (or degeneracy) factor which depends on the model under consideration (EPRL/FK or BC), and the energy is now given by $E'(\rho,\nu)=E(\nu)-a\rho/2$.

\section{Partition function in the self-dual theory}
\label{sec:4}

\noindent In \cite{FGNP} we suggested that the microstates of a black hole in the self-dual quantum theory are described by the analytically-continued dimension of the $\SU(2)$ Chern--Simons Hilbert space. We are now going to show that this proposal does also describe the near-horizon radiation, which bridges the gap with the approach to thermality presented in the previous section.

In LQG, black holes can be described in terms of an $\SU(2)$ Chern--Simons theory, where the coupling constant $k$ is proportional to the horizon area $A$, and the underlying spacetime manifold has the topology $\mathbb{S}^2\times\mathbb{I}$, where $\mathbb{S}^2$ is a $p$-punctured two-sphere and $\mathbb{I}$ a segment in $\mathbb{R}$. The punctures come from spin network edges crossing the horizon, and carry quanta of area. The black hole microstates are invariant tensors (intertwiners) in the tensor product $\bigotimes_e j_e$ of the representations $j_e$ coloring the edges crossing the horizon. For a fixed set $\vec{\jmath}=(j_1,\dots,j_p)$ of representations, the microstates span a finite-dimensional Hilbert space whose dimension can be computed with tools from quantum groups and is given by \cite{KMdim}
\be\label{su(2) dim}
\mathcal{D}(\vec{\jmath})=\f{2}{k+2}\sum_{d=1}^{k+1}\sin^2\left(\f{\pi d}{k+2}\right)\prod_{e=1}^p\f{\ds\sin\left(\f{\pi d}{k+2}d_e\right)}{\ds\sin\left(\f{\pi d}{k+2}\right)},
\ee
where $d_e=2j_e+1$ is the dimension of the spin $j_e$ representation of $\text{U}_q(\su(2))$.

In the standard microcanonical derivation of entropy in LQG, the dimension $\mathcal{D}(\vec{\jmath})$ is summed over all possible number of punctures and spin values labelling these (distinguishable) punctures, together with a constraint enforcing that the macroscopic horizon area $A$ be the sum of the microscopic contributions $A_e=8\pi\gamma\lp^2\sqrt{j_e(j_e+1)}$. The partition function defined in this way reproduces the Bekenstein--Hawking formula (and its logarithmic corrections) once $\gamma$ is fixed to a particular real value, and one can furthermore show that the dominant contribution to the entroy comes from small spins. As argued in the introduction, this philosophy goes against the logic according to which the coupling constants should run and take effective values in the IR, which is the relevant regime for comparing an entropy calculation with the semiclassical Bekenstein--Hawking formula.

Our proposal for working with $\gamma=\i$ can evidently not be applied in the above-described picture (it would lead to an imaginary entropy). To make sense out of it, one has to modify the structure of the quantum theory itself. There seem to be two consistent and mutually compatible ways of doing so (see \cite{Muxin} for an alternative proposal that postulates a canonical partition function).
\begin{flist}
\vspace{-0.1cm}
\item[$a)$] \textit{Analytic continuation of the Chern--Simons level.}\vspace{0.1cm}\\
Although the precise proportionality coefficient between the level $k$ and the horizon area $A$ has been debated, it is possible to write the Chern--Simons theory on the horizon in such a way that $k\propto A/\gamma$. Under the assumptions that $k$ (i.e. the area) is large, and that in the pre-factor and in the upper bound of the sum in \eqref{su(2) dim} one can replace $k$ by its modulus $|k|$, the level in the self-dual theory with $\gamma=\i$ becomes $k=\i\lambda$ and \eqref{su(2) dim} becomes
\be\label{Zjl}
\mathcal{Z}(\vec{\jmath})=\f{2}{\lambda}\sum_{d=1}^\lambda\sinh^2\left(\f{\pi d}{\lambda}\right)\prod_{e=1}^p\f{\ds\sinh\left(\f{\pi d}{\lambda}d_e\right)}{\ds\sinh\left(\f{\pi d}{\lambda}\right)}.
\ee
In fact, regardless of wether $k$ depends on $\gamma$ or not, working with a purely imaginary level can be understood as describing one chiral component of $\SL(2,\mathbb{C})$ Chern--Simons theory \cite{WittenCS}. The apparent drawback of this proposal is that is requires to set $\gamma=|\gamma|$ in order to ensure that the area spectrum remains real. However, this is an immediate consequence of the fact that we are analytically continuing the spectrum constructed from the $\SU(2)$ theory, and we expect that a proper understanding of the geometrical operators in the self-dual quantum theory will raise this ambiguity.
\vspace{-0.1cm}
\item[$b)$] \textit{Analytic continuation of the} $\SU(2)$ \textit{spins.}\vspace{0.1cm}\\
If we think of the punctures as coming from the bulk, where the quantum theory is defined in terms of a (so far unknown) self-dual theory, there is a priori no reason for them to carry finite-dimensional half-integer spin representations. There are however two natural requirements that can be imposed on the type of representations coloring an $\SL(2,\mathbb{C})$ spin network state. The first one is that they be (anti) self-dual, in the sense that $\vec{L}\mp\i\vec{K}=\vec{0}=\vec{J}^\mp$ as an equation for operators acting on states $|(\rho,k),(j,n)\ra$, and the second one is that the area operator be real (i.e. satisfy the area reality condition). When expanding the representations of $\SL(2,\mathbb{C})$ in terms of UIR's of the subgroup $\SU(1,1)$, it appears that these two conditions can be met if one considers the representation labels $(\rho,k)=(-\i(j+1),\pm j)$ or $(\pm\i j,-(j+1))$, where the spin $j=\i s-1/2$ labels elements of the continuous series ($s\in\mathbb{R}$). At the level of the dimension \eqref{su(2) dim}, the substitution $j_e\rightarrow\i s_e-1/2$ leads to
\be\nn
\mathcal{Z}(\vec{s})=\f{2}{k+2}\sum_{d=1}^{k+1}\sin^2\left(\f{\pi d}{k+2}\right)\prod_{e=1}^p\f{\ds\sinh\left(\f{2\pi d}{k+2}s_e\right)}{\ds\sin\left(\f{\pi d}{k+2}\right)}.
\ee
One can immediately note the similarity between this expression and \eqref{Zjl}.
\end{flist}
\vspace{-0.1cm}

$\mathcal{Z}(\vec{\jmath})$ and $\mathcal{Z}(\vec{s})$ are the two candidate partition functions for describing the black hole in the self-dual theory. In the limit in which $j_e\rightarrow\infty$ (or $s_e\rightarrow\infty$) and $\lp\rightarrow0$ with $j_e\lp$ finite, their behavior is the same and given by the last term in the sum. Focusing for example on \eqref{Zjl}, one can write that
\be\nn
\mathcal{Z}(\vec{\jmath})\approx\mathcal{Z}_0=\f{2\sinh^2\pi}{\lambda}\prod_{e=1}^p\f{\sinh(\pi d_e)}{\sinh\pi},
\ee
which immediately leads to
\be\nn
\log\mathcal{Z}(\vec{\jmath})\approx\log\mathcal{Z}_0\approx2\pi\sum_{e=1}^pj_e=\f{A}{4\lp^2}.
\ee
The same asymptotic behavior holds for $\mathcal{Z}(\vec{s})$.

Now, a straightforward calculation shows that one can write
\be\nn
\mathcal{Z}_{0}=\f{2\sinh^2\pi}{\lambda}\prod_{e=1}^p\mathcal{Z}_e,\qquad\text{with}\qquad\mathcal{Z}_e=\sum_{n=0}^{2j_e}e^{-\beta E_e(n)},
\ee
which justifies the interpretation of the analytically-continued Chern--Simons Hilbert space dimension as a partition function. Here $\beta=2\pi/a$ is the Unruh temperature, and $E_e(n)=an+E_e(0)=a(n-j_e)$ is a discrete and finite energy spectrum with negative lowest energy $E_e(0)=-aj_e$. We therefore see that the partition function $\mathcal{Z}_0$ can be written as the product of $p$ different partition functions $\mathcal{Z}_e$, each corresponding to a system of particles at equilibrium at Unruh temperature and with discrete energy levels.

To interpret this result, one has to remember that the punctures carry spin labels $j_e$ determining the areas $A_e$ of the facets $f_e$ that tesselate the horizon. In fact, a puncture carries not only an area element, but the whole Hilbert space $V_{j_e}$ associated to the representation of spin $j_e$, a basis of which is labelled by the magnetic numbers $n=-j_e,\dots,j_e$. The significance of the above expression for $\mathcal{Z}_0$ is therefore that a stationary observer at distance $a^{-1}$ from a facet $f_e$ associates to the states labelled by $n$ an energy $E_e(n)=a(n-j_e)$. Furthermore, one sees that the variation of energy measured by the observer is
\be\nn
\delta E=\f{\delta A}{8\pi G}a,
\ee
which coincides with the Frodden--Gosh--Perez energy \cite{FGP} that was derived from the boost Hamiltonian by Bianchi \cite{BianchiBH}.

The lowest energy level $E_e(0)=-aj_e$, which is negative and becomes very low in the large spin limit, is in fact responsible for the entropy, while the excited modes contribute less. One can furthermore write the partition function in terms of a degeneracy factor given by the lowest energy as follows:
\be\nn
\mathcal{Z}_e=\sum_{n=0}^{2j_e}e^{-\beta E_e(n)}=\sum_{n=0}^{2j_e}g_ee^{-\beta an},
\ee
where the degeneracy factor is holographic and given by $g_e=\exp(2\pi j_e)=\exp(A_e/4\lp^2)$. As a consequence, one can interpret the quantum system described by this partition function as a degenerate system whose energy spectrum is given by $E'(n)=an$, and where each energy level is degenerate.

It is very interesting to note that we recover here the assumptions that were made in \cite{GNP} order to motivate the introduction of canonical and grand canonical partition functions for describing black hole thermodynamics with inputs from the self-dual theory. These ingredients are that the punctures have to be treated as being indistinguishable (either with a Gibbs factor or with a proper bosonic or fermionic statistics), and that they carry a holographic degeneracy given by $g_e$. As shown in \cite{GNP}, these inputs have far reaching consequences, and imply in particular that the number of punctures scales like $p\propto\sqrt{A}/\lp$. This can be understood as a ``continuum limit'' in the sense that the state is fine grained in spite of being described by large quantum numbers.

\section{Conclusion}
\label{sec:5}

\noindent In this short paper, we have argued that a self-dual formulation of LQG might indeed be the proper framework in which to describe black hole thermodynamics, and provided, in addition to the original observation \cite{FGNP}, two new evidences supporting this fact. The first one is that the derivation of the near-horizon radiation proposed by Bianchi in \cite{BianchiBH} can also be understood in terms of a decomposition of spin foam states into self-dual and anti self-dual parts. Indeed, we have shown that the density matrix obtained with this decomposition describes a thermal state once it is traced over one chiral component. The second fact is that the analytically-continued dimension of the Chern--Simons Hilbert space can be written in such a way that it assigns to each puncture an Unruh temperature $T=1/\beta$, a discrete energy spectrum, and a holographic degeneracy. It is interesting to note that the fact that the partition function of the self-dual theory might be thermal and related to Chern--Simons theory was already pointed out in \cite{Smolin}.

The description proposed in section \ref{sec:3} resembles closely the derivation of the thermality of a Rindler wedge. However, instead of taking place in spacetime, our description relies on the decomposition of the tangent Lorentzian space into (anti) self-dual parts. This is quite natural since the boost Hamiltonian in \eqref{hamiltonian} does indeed live in the tangent space. It would be interesting to understand in more details the relationship between these two approaches.

There is an increasing amount of evidence supporting the relevance of the description of black hole entropy from the point of view of self-dual quantum gravity \cite{FGNP,Smolin,BTZ,KMS,CMC,BN,Muxin}. We hope that in the near future this framework will be clarified and developed more thoroughly, and that these results will serve as a starting point for a systematic investigation of the self-dual structure of LQG and spin foams.

\section*{Acknowledgments}

\noindent We would like to thank Alejandro Perez for collaboration at the early stages of this work, Wolfgang Wieland and Lee Smolin for discussions and comments, and Jeff Hnybida for bringing \cite{Huszar} to our attention. MG is supported by the NSF Grant PHY-1205388 and the Eberly research funds of The Pennsylvania State University.

\appendix

\section{Representation theory of $\SL\boldsymbol{(2,\mathbb{C})}$}
\label{appendix1}

\noindent Let us denote by $\vec{L}=(L_i)_{i=1,2,3}$ and $\vec{K}$ the generators of infinitesimal rotations and boosts. They generate the Lie algebra $\sll(2,\mathbb{C})$, and their commutation relations are given by
\ba
&[L_i,L_j]=\eps_{ij}{}^kL_k,\qquad[K_i,K_j]=-\eps_{ij}{}^kL_k,&\nn\\
&[K_i,L_j]=\eps_{ij}{}^k K_k,&\nn
\ea
where $\eps_{ijk}$ is the totally antisymmetric tensor, and the indices are lowered and raised with the flat Euclidean metric $\delta_{ij}$.

Unitary irreducible representations of $\SL(2,\mathbb{C})$ are labelled by a couple $(\rho,k)$ where $\rho$ is a real number and $k$ is an integer. These representations are infinite dimensional, and the associated Hilbert space is a Verma module $V_{\rho,k}$. The quadratic Casimir operators are given by $C_1=\vec{L}^2-\vec{K}^2=k^2-\rho^2-1$ and $C_2=\vec{L}\cdot\vec{K}=k\rho$.

The space $V_{\rho,k}$ can be decomposed into $\SU(2)$ modules $V_j$ (where $j$ is a half-integer spin) according to the Iwazawa decomposition of the Lorentz group given by $V_{\rho,k}=\oplus_{j=k/2}^\infty V_j$. With this decomposition, basis vectors are denoted by $|(\rho,k),(j,n)\ra$ or simply $|j,n\ra$, and diagonalize the Casimir $\vec{L}^2=\vec{L}\cdot\vec{L}=L_iL_j\delta^{ij}$ and the third components $L_3$ of the rotations in the following way:
\be\nn
\vec{L}^2|j,n\ra=j(j+1)|j,n\ra,\qquad L_3|j,n\ra=n|j,n\ra,
\ee
with $j\in\mathbb{N}/2$ and $n=-j,\dots,j$.

\section{Overlap coefficient between the two bases}
\label{appendix2}

\noindent The generic expression for the overlap coefficient between the two bases is given by \cite{Huszar}
\begin{widetext}
\ba
\la\mu,\nu|j,n\ra
&=&\delta_{\mu,n}\sqrt{\f{2j+1}{4\pi}}C_{k-\mu}(\ell_\pm,m_\pm)\f{\Gamma(\ell'-m'+1) \Gamma(\ell'+\mu'+1)}{\Gamma(k'-\mu'+1)\Gamma(k'+\i\rho+1)}\left(\f{\Gamma(j+k'+1)\Gamma(j-\mu'+1)}{\Gamma(j-k'+1)\Gamma(j+\mu'+1)}\right)^{1/2}\nn\\
&&\times\left(\f{\overline{\Gamma(-\ell_++\ell_-+j+1){\Gamma(\ell_+-m_++1)}\Gamma(\ell_++m_++1)}}{\overline{\Gamma(\ell_+-\ell_-+j+1)}\overline{\Gamma(\ell_--m_-+1)}\overline{\Gamma(\ell_-+m_-+1)}}\right)^{1/2}
{}_3F_{2}\left[\begin{array}{c}
-j+k'\;;\;j+k'+1\;;\;\ell'-m'+1\\
k'-\mu'+1\;;\;k'+\i\rho+1
\end{array}\right].\nn
\ea
\end{widetext}
This expression is written in terms of the generalized hypergeometric function for a unit argument $z=1$, i.e.
\be\nn
{}_3F_{2}\left[\begin{array}{c}a_1\;;\;a_2\;;\;a_3\\b_1\;;\;b_2\end{array}\right]={}_3F_{2}\left[\begin{array}{c}a_1\;;\;a_2\;;\;a_3\\b_1\;;\;b_2\end{array}\;;\;z=1\right],
\ee
the Euler Gamma function $\Gamma(z)$ defined for $z\in\mathbb{C}$, and a function $C_{k-\mu}(\ell_\pm,m_\pm)$ which depends on the sign of $k-\mu$ in the following way:
\be
C_{k-\mu}(\ell_\pm,m_\pm)=
\left\{
\begin{array}{l}
1
\phantom{\f{\ds\sin\pi(\ell_+-m_+)}{\ds\sin\pi(\ell_--m_-)}\quad}
\text{if } k\leq\mu,\nn\\
\f{\ds\sin\pi(\ell_+-m_+)}{\ds\sin\pi(\ell_--m_-)}
\phantom{1\quad}
\text{if } k\geq\mu.
\end{array}
\right.
\ee
Finally, following \cite{Huszar}, we have introduced the following notations in order to lighten the formula for the overlap:
\ba
&\ds k'=\f{1}{2}(|k+\mu|+|k-\mu|),\qquad\mu'=\f{1}{2}(|k+\mu|-|k-\mu|),&\nn\\
&\ds\ell'=\f{1}{2}(k'+\i\rho-1),\qquad m'=\f{1}{2}(\mu'+\i\nu).&\nn
\ea

\section{Overlap coefficient for the Barrett--Crane model}
\label{appendix3}

\noindent In the case of the Barrett--Crane model, the simple representations of the Lorentz group defining the states are of the form $|(\rho,0),(0,0)\ra$, which we denote simply by $|0(\rho)\ra$. A simple calculation then shows that the square of the overlap coefficient is given by
\ba\nn
\big|\la\mu,\nu|0(\rho)\ra\big|^2=\delta_{\mu,0}\f{1}{2\rho}\f{\sinh(\pi\rho)}{\cosh(\pi\rho)+\cosh(\pi\nu)} .
\ea
Up to the rational function $Q_j(\rho,\nu)$, this expression is similar to the one obtained for the EPRL/FK model when $j$ is an integer.

\section{Asymptotics of the overlap for any state}
\label{appendix4}

\noindent We have seen that the overlap coefficients in the case of the EPRL/FK and BC models have quite similar expressions. This is responsible for the fact that the reduced density matrix computed by tracing over the anti self-dual part exhibits a thermal behavior.

A natural question to ask is wether this is a generic feature of the overlap coefficients. One can see that this is indeed the case by looking at the large $\nu$ behavior of the overlap coefficients. It was shown in \cite{Huszar} that when $\nu\gg1$ is real the following asymptotic formula holds:
\be\nn
\la\mu,\nu|j,n\ra=\delta_{\mu,n}\Psi_{\rho,k}(j,n)\left(\f{\i\nu}{2}\right)^\alpha e^{-\pi|\nu|/2}\big(1+\circ(1)\big),
\ee
where $\alpha=(2\ell_++1+j-k)$. Here $\Psi_{\rho,k}(j,n)$ does not depend on $\nu$ and is given by
\begin{widetext}
\be
\Psi_{\rho,k}(j,n)=\f{\sqrt{(2j+1)\pi}\ \Gamma(\ell_++\ell_-+j+2)\Gamma(j-n+1)}{\Gamma(2\ell_++2)\Gamma(\ell_++\ell_--n+2)\Gamma(-\ell_+-\ell_-+j) \Gamma(j +n+1)}
\ {}_2F_{2}\left[ 
\begin{array}{c}
\ell_++\ell_--j+1\;;\;\ell_++\ell_-+j+2\nn\\
2\ell_++2\;;\;\ell_++\ell_--n+2
\end{array}
\right],
\ee
\end{widetext}
where $\ell_\pm=(k\pm\i\rho-1)/2$ are the complex spins associated to the representation $(\rho,k)$, and ${}_2F_{2}$ is the generalized hypergeometric series for a unit argument.

\end{document}